\documentclass[twocolumn,showpacs,superscriptaddress,aps,prl]{revtex4}

\usepackage{graphicx}
\usepackage{amssymb,amsmath}
\usepackage{dcolumn}
\usepackage{bm}

\begin{document}

\preprint{$\#$1 sakiyama}

\title{Short-range Charge and Spin Superstructures in Doped Layered Co Perovskites}

 \author{N. Sakiyama}
 \affiliation{Neutron Science Laboratory, Institute for Solid State
Physics, University of Tokyo, Tokai 319-1106, Japan}
 \author{I. A. Zaliznyak}
 \email[Address correspondence to ]{zaliznyak@bnl.gov}
 \affiliation{DCMPMS,
 Brookhaven National Laboratory, Upton, NY 11973}
 \author{S. -H. Lee}
 \affiliation{Department of Physics, University of Virginia,
Charlottesville, Virginia 22904, USA}
 \author{Y. Mitsui}
 \author{H. Yoshizawa}
 \affiliation{Neutron Science Laboratory, Institute for Solid State
Physics, University of Tokyo, Tokai 319-1106, Japan}

\date{\today}

\begin{abstract}

We have investigated cobaltite relatives of the layered perovskite
cuprates and nickelates, Pr$_{2-x}$Ca$_x$CoO$_4$ ($0.39 \leq x \leq
0.73$) and La$_{2-x}$Sr$_x$CoO$_4$ ($x = 0.61$), using elastic
neutron scattering. We have discovered doping-dependent
incommensurate short-range ordering of charges and magnetic moments,
which in cobaltites occur in the range of heavy doping, $ 0.5
\lesssim x \lesssim 0.75$. The charge order exists already at room
temperature and shows no change on cooling. The incommensurability
of its propagation vector, ${\bf Q}_c = (\epsilon_c,0,l)$, roughly
scales with the concentration of Co$^{2+}$ ions, $\epsilon_c \sim
(1-x)$. Magnetic order is only established at low T$\lesssim 40$ K
and has twice larger periodicity, indicating a dominant
antiferromagnetic correlation between the nearest Co$^{2+}$ spins.

\end{abstract}


\maketitle



Physical origins of the ubiquitous doping-dependent incommensurate
charge and spin ordering (CO and SO) in doped
La$_{2-x}$Sr$_{x}$CuO$_{4}$ cuprates and their relevance to
mechanisms of the high-temperature superconductivity have been
subjects of intense research for more than a decade but still remain
a mystery. Numerous experimental studies have shown, that holes in
weakly doped cuprates exhibit an in-plane CO, whose propagation
vector scales roughly proportionally with doping $x$ and magnetic
order, whose modulation period is twice larger
\cite{LNSCuO_stripe_nature,LSCuO_yamada_plot,Kivelson_RMP2003}.
Similar findings were reported in closely related, isostructural but
insulating layered perovskite nickelates La$_{2-x}$Sr$_{x}$NiO$_{4}$
\cite{ChenCheongCooper_PRL1993,LNiOd_stripe,Yoshizawa_PRB2000_LSNO_stripe_step,
Kajimoto_PRB2003_LSNiO_stripe_PD,LSNiO_stripe_SHL}. A simple picture
of the simultaneous real-space static ordering of charges and spins
yielding such CO and SO is provided by the charge stripe model,
where doped holes segregate into lines separating stripes of
antiferromagnetically ordered domains \cite{LNSCuO_stripe_nature}.

The charge stripe picture is strongly supported by theoretical
analysis of the two-dimensional (2D) Hubbard model, which is
believed to describe high-T$_c$ cuprates
\cite{Kivelson_RMP2003,Zaanen1989,Emery1993,stripe_theory,mfa_theory}.
In the 2D Hubbard model charge stripe and antiferromagnetic spin
order are intimately coupled, as both are signatures of the same
ordering instability of interacting itinerant charges. This model,
however, discards charge immobilization effects, such as their
trapping by local polarons. Doped holes in charge stripe picture are
intrinsically highly mobile and decrease their energy by
one-dimensional (1D) delocalization. While this assumption seems
justified for moderately to highly doped cuprates, which show
metallic conductivity and where an activated charge transport occurs
only at low temperatures and is associated with the onset of CO in
the form of "parallel stripes" aligned with Cu-O bonds
\cite{Ichikawa_PRL2000}, it is in question for lightly doped
insulating cuprates and for the insulating nickelates
\cite{Anisimov_PRL1992,Zaanen_PRB1994}, where CO corresponding to
the so-called "diagonal" stripes with charge lines at 45$^\circ$ to
Cu-O bonds is observed.

In fact, cuprates are rather exceptional among Mott-Hubbard
insulators (MHI), as most MHI cannot be made metallic by doping:
strong polaronic self-localization of doped holes hinders metallic
transport \cite{Zaanen_PRB1994}. Theoretically, however, stripe-like
superstructures whose period depends on the doping level are also
predicted in systems with localized doped charges
\cite{KhomskiiKugel}. They are natural response of the crystal
lattice to a strain associated with doping and can be explained by
considering the system's elastic energy. In this picture CO is a
cooperative ordering of polarons driven by lattice elastic
interactions \cite{Zaanen_PRB1994,KhomskiiKugel}. So, are CO and SO
scattering patterns with spin and charge incommensurabilities
related by $\delta_s \sim \epsilon_c/2$ a characteristic signature
of charge stripes -- a 1D segregation of itinerant charges, or a
ubiquitous feature of doped MHI, independent of conducting
properties?

In an attempt to answer this question we have studied layered
perovskite cobalt oxide series Pr$_{2-x}$Ca$_x$CoO$_4$ (PCCoO),
$0.39 \leq x \leq 0.73$, and La$_{2-x}$Sr$_x$CoO$_4$ (LSCoO), $x =
0.61$, isostructural with 2D cuprates and nickelates.
The undoped ($x=0$) LSCoO parent material is a charge-transfer (CT)
antiferromagnetic insulator with T$_N \approx 275$ K
\cite{Yamada_PRB1989}. Magnetic Co$^{2+}$ are in 3$d^7$
($t_{2g}^5e_g^2$) state and, like Cu$^{2+}$, are Kramers ions,
albeit with spin S=3/2. Hund's coupling in cobaltites is in close
competition with the crystal field (CF), so doped holes can yield
Co$^{3+}$ (3$d^6$) in an intermediate (IS, S=1) or low spin (LS,
S=0) state \cite{Zaliznyak,Moritomo_PRB1997,SpinStateNote}. In
either case the ground state in the crystal field is $S^z = 0$
singlet and is effectively non-magnetic at low T \cite{Zaliznyak}.
This is a fundamental consequence of Kramers time-reversal symmetry
and makes comparison with cuprates particularly meaningful (Hund's
rule is not at play for a single hole in $3d^9$ Cu$^{2+}$, while
covalency and strong correlation yield a non-Hund S = 0 singlet
$3d^8$ state for doped holes). Similar to the cuprate case, magnetic
ordering temperature in LSCoO drops dramatically with doping. For $x
= 0.5$ glassy SO only appears below $\sim 30$ K.

Unlike cuprates, layered LSCoO cobaltites remain insulating
throughout the doping range $0 \leq x \lesssim 1$, which is in fact
typical for a doped CT/MHI
\cite{Moritomo_PRB1997,Matsuura_JPCS1988}. Doped holes in cobaltites
are strongly localized, so that $ab$-plane resistivity in LSCoO is
in the $\sim 10$ $\Omega \cdot $cm ($x = 1$) to $\gtrsim 10^4$
$\Omega \cdot $cm ($x \lesssim 0.5$) range and shows polaronic
activated behavior with activation energy $E_a \sim 1500$ K to $E_a
\gtrsim 5000$ K, respectively \cite{Moritomo_PRB1997}. Short-range
checkerboard charge order (CCO), which has been observed by neutron
scattering in the $x=0.5$ sample at $T \lesssim 825$ K
\cite{Zaliznyak} is thus a correlated polaron glass phase. It should
be noted, that CCO can be considered a limiting case of charge
stripe order with shortest possible stripe spacing, and this is how
it was interpreted in the $x = 0.5$ nickelate
\cite{Yoshizawa_PRB2000_LSNO_stripe_step,Kajimoto_PRB2003_LSNiO_stripe_PD}.
In LSCoO, however, detailed investigation of CO diffuse scattering
showed no evidence for incipient 1D charge stripes
\cite{Savici_PRB2007}. In addition, polaron CO is totally
independent of spins, which only order at $T \lesssim 30$ K.

\begin{figure}[!t]
\begin{center}
\includegraphics[width=\linewidth]{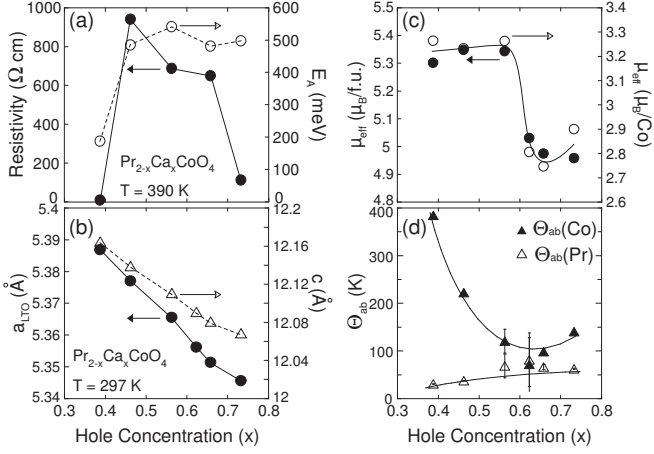}
\caption{Doping dependence of electrical, structural and magnetic
properties of PCCoO. (a) resistivity at the highest measured
temperature (closed symbols) and its activation energy (open). (b)
F4/mmm ("LTO") lattice parameters refined at room temperature by
powder Xray. (c) effective magnetic moment $\mu_{eff}$ per formula
unit obtained by one-component Curie-Weiss analysis of the in-plane
magnetic susceptibility $\chi_{ab}$(T), 50 K $\leq$ T $\leq$ 300 K,
which includes Pr$^{3+}$ contribution (left scale, closed symbols).
Open symbols are $\mu_{eff}$ per Co ion obtained from two-component
Curie-Weiss fit with fixed $\mu_{eff}^{Pr} = 3.58 \mu_B$; (d) shows
Weiss temperatures for Co (filled) and Pr (open) from this fit,
lines are guide for the eye.
\label{Fig0_xdep} }
\end{center}
\vspace{-0.4in}
\end{figure}

Polaronic self-localization of doped holes is similarly strong in
PCCoO series. Electrical resistibility and thermal activation energy
measured by PPMS in our single crystal samples are shown in Figure
\ref{Fig0_xdep}(a). Samples were grown by the floating zone method
and their uniformity was verified by Scanning Electron Microscopy.
The actual chemical composition $x$ was then determined using the
Inductively Coupled Plasma method. The absence of impurity phases
was confirmed by powder x-ray diffraction. Samples were further
characterized by measuring static magnetic susceptibility by MPMS
SQUID magnetometer. Typical data is presented in the inset to Figure
\ref{Fig3_PD-mu_int} (b). Weiss temperature $\Theta_W$ and Co
effective magnetic moment resulting from the Curie-Weiss analysis of
its T-dependence are shown in Figure \ref{Fig0_xdep}(d) and (c).

For neutron diffraction studies we used $\approx 20$ mm long, $\phi
\approx 6$ mm cylindrical single crystal pieces of m $\approx 2$ g.
Mosaic of the fundamental Bragg reflections was $\lesssim 0.5
^\circ$, which is consistent with the instrumental resolution and
shows high crystalline quality. Neutron experiments were performed
on 4G and 5G 3-axis thermal neutron spectrometers at the JRR-3 at
JAEA, Tokai, Japan, with $40'-40'-40'-40'$ beam collimation and BT9
spectrometer at the NIST Center for Neutron Research, with beam
collimation $\approx 40'-47'-44'-80'$(open). Wave vector of the
incident and scattered neutrons was selected by the (002) Bragg
reflection from pyrolitic graphite (PG) and fixed at $k_i \simeq
2.67 \text{\AA}$. Contamination from higher order reflections was
suppressed by $1'' - 2''$ thick PG transmission filters. Sample was
mounted in a closed-cycle He gas refrigerator ($T \ge 0.7$ K) with
$a-b$ plane vertical and $(h0l)$ reciprocal lattice zone in the
horizontal scattering plane. We index wave vectors in the F4/mmm
lattice with the unit cell $\sqrt{2} a \times \sqrt{2} a \times c$
compared to the I4/mmm high-temperature tetragonal (HTT) lattice of
cuprates with $a \approx 3.8 \AA$. We assign "LTO" index to the
in-plane lattice parameters thus defined, Figure \ref{Fig0_xdep}(b),
since our samples are actually in the low-temperature orthorombic
(LTO) phase, but the orthorombic distortion is way too small to be
resolved in the present measurements \cite{Savici_PRB2007}.

\begin{figure}[!t]
\begin{center}
\includegraphics[width=\linewidth]{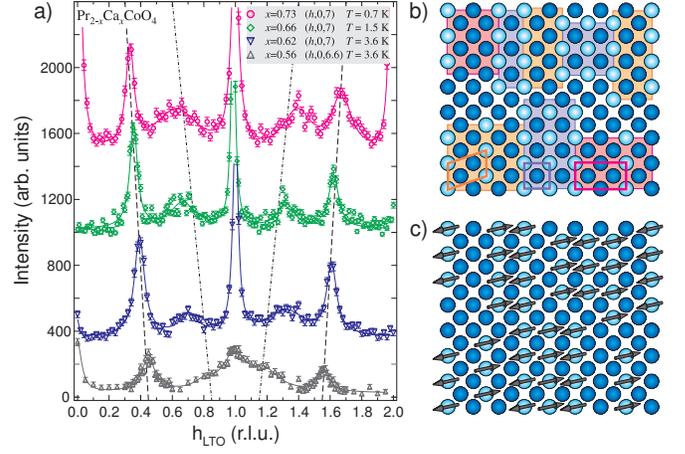}
\caption{(a) Elastic scans along $(h,0,7)$ or $(h,0,6.6)$, showing
charge and spin order peaks at ${\bf Q}_c = (2n \pm \epsilon_c,0,l)$
and ${\bf Q}_s = (2n \pm \delta_s ,0,l)$, respectively. Solid lines
are Lorentzian fits; broken lines trace $\epsilon_c \sim 2
\delta_s$. Peak at $(007)$ contains superlattice Bragg intensity
resulting from the LTO distortion but is contaminated by lattice
Bragg scattering of $\lambda/2$ neutrons. (b) sketch of glassy
charge-ordered state with a mixture of different commensurate
superlattice fragments accommodating the nominal doping $x \approx
0.7$; (c) antiferromagnetic correlations governed by
Co$^{2+}-$Co$^{3+}-...-$Co$^{2+}$ superexchange between Co$^{2+}$
spins (arrows) developing in this CO state. \label{Fig1_COSOscans} }
\end{center}
\vspace{-0.4in}
\end{figure}

Typical low-temperature scans along $h_{LTO}$ direction, revealing
superlattice scattering in several PCCoO samples are shown in Figure
\ref{Fig1_COSOscans}. Two sets of peaks whose position varies
roughly linearly with doping can be identified. Sharper and more
intense peaks at $\bm{Q}_s = \bm{G} \pm (\delta_s,0,l)$, where
$\bm{G}$ is the wave vector of a fundamental lattice Bragg
reflection, quickly disappear upon heating to T $\gtrsim 40$ K and
decrease in intensity with increasing $Q_s$ for different $\bm{G}$,
roughly following the square of the Co$^{2+}$ magnetic form factor.
Hence, we identify these peaks as magnetic scattering arising from
spin order. Their position varies roughly linearly with the
remaining Co$^{2+}$ content, $\delta_s(x) \sim (1-x)$. Weak diffuse
peaks at $\bm{Q}_c = \bm{G} \pm (\epsilon_c,0,l)$, whose position
varies roughly as $\epsilon_c(x) \sim 2(1-x)$, are much broader and
in fact increase in intensity with increasing $Q_c$ in different
$\bm{G}$ zones, roughly $\sim Q_c^2$. We therefore identify them as
superstructural scattering arising from atomic displacements
accompanying short-range charge/valence order. These peaks remain
unchanged both in width and intensity upon heating to T $\approx
300$ K.

Strong peak at $(1,0,7)$ is consistent with Bragg scattering arising
from the LTO lattice distortion. Its intensity grows rather
gradually below $\approx 300$ K and then usually decreases below $T
\sim 100$ K. Similar peak is present at other integer $l$, in
particular at $l = 6$. This type of reflections was observed by
neutron diffraction in the undoped La$_2$CoO$_4$
\cite{Yamada_PRB1989}. As $x \rightarrow 0.5$, $\epsilon_c(x)
\rightarrow 1$ and LTO Bragg scattering at $h = 1$ overlaps with
weak CO peak at $\bm{G \pm Q}_c$. The latter, however, persist to
much higher temperature, T $\sim 700 - 800$ K, and is much broader.
The lower scan in Fig. \ref{Fig1_COSOscans}(a) for $x = 0.56$
performed at $l = 6.6$ has only weak Bragg contribution at $h = 1$,
while CO scattering is practically unchanged compared to $l = 7$.

The $l-$dependence of CO scattering measured in our PCCoO samples
shows no peaks due to inter-plane coherence ($\xi_c^l \lesssim 0.2
c$), but smooth intensity modulation resembling that reported for
$x=0.5$ LSCoO \cite{Zaliznyak}. This indicates that scattering
arises from similar distortions of CoO$_6$ octahedra, which are
associated with charge polarons. In $x = 0.46$ and $0.39$ PCCoO
samples CO peak is at $h = 1$, indicating a finite region of
stability of the CCO at $x \leq 0.5$. The in-plane peak width in
these samples is comparable to that in $x = 0.5$ LSCoO,
corresponding to CCO correlation length $\xi_c^h$ of $3 - 5$ LTO
lattice spacings. For $x \geq 0.5$, where polaron correlations are
incommensurate, CO peaks are at least twice broader, corresponding
to $\xi_c^h \lesssim 10 \AA$. Despite different tolerance factors
and band-widths compared to PCCoO, our $x=0.61$ LSCoO sample shows
very similar glassy CO and SO correlations, which folow the same
doping trend.

The picture of CO at $x \geq 0.5$ arising from these observations is
illustrated in Figure \ref{Fig1_COSOscans}(b). Polarons associated
with Co$^{2+}$ sites, whose concentration is $n_e = 1-x$, build up
patches of commensurate superlattices corresponding to $n_e =
1/2,1/3,1/4, etc.$, where particular $n_e$ is locally favored by
La/Sr doping fluctuation. Their long-range coherence, however, is
frustrated by the charge neutrality condition imposed by the average
Sr$^{2+}$ concentration $x$. Hence, glassy CO state results, which
is made of a mixture of $\sim 1$ nm sized domains of commensurate
polaron superlattices, accommodating the average charge doping.
Consequently, CO scattering has an average incommensurability of
$\epsilon_c(x) \approx 2(1-x)$ .

\begin{figure}[!t]
\begin{center}
\includegraphics[width=0.6\linewidth]{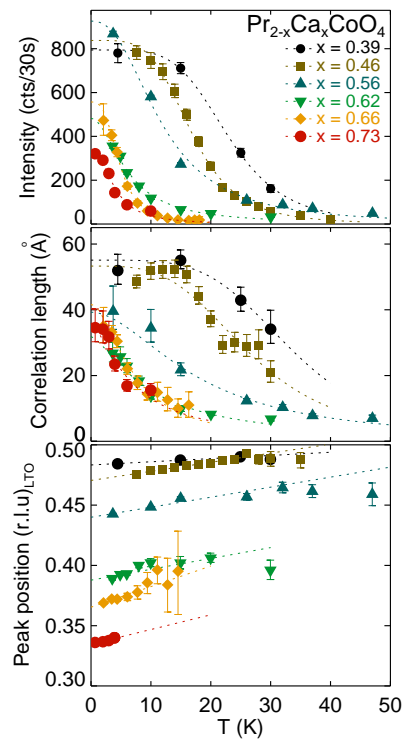}
\caption{Temperature dependence of magnetic SO peak parameters in
PCCoO. (a) peak intensity cross-normalized to $\approx 30$ sec BT9
monitor count for equal sample mass; (b) in-plane correlation length
$\xi_{ab}$ (inverse half-width at half maximum (HWHM) of Lorentzian
fit); (c) peak position. \label{Fig2_SO_Tdep} }
\end{center}
\vspace{-0.4in}
\end{figure}

Magnetic SO scattering with roughly half the CO incommensurability
simply corresponds to the antiferromagnetic order of Co$^{2+}$
spins, Figure \ref{Fig1_COSOscans}(c). Magnetic correlations are
apparently governed by Co$^{2+}-$Co$^{3+}-...-$Co$^{2+}$
superexchange (note that $\Theta_W > 0$ for all $x$, Fig.
\ref{Fig0_xdep}(d)). They decrease with increasing number of
intervening Co$^{3+}$, but also extend between the different CO
superlattice domains, which explains why the in-plane SO correlation
length $\xi_s^h$ is 3-4 times larger than $\xi_c^h$. This also
explains the decrease with $x$ of the SO temperature seen in panel
(a) of Figure \ref{Fig2_SO_Tdep}, which shows the T-dependence of
the SO peak parameters. Typical of a glassy spin freezing, SO peak
intensity increases smoothly with decreasling T, while the
correlation length $\xi_s^h$, Figure \ref{Fig2_SO_Tdep}(b),
increases and seems to saturate roughly where the intensity reaches
half its maximum. The temperature dependence of the SO peak
position, Figure \ref{Fig2_SO_Tdep}(b), also shows an interesting
trend, which can be simply understood in this model. Indeed, as
longer-periodic CO polaron superlattices present in a given sample
gradually develop spin correlations with decreasing T, the average
magnetic modulation period becomes longer and $\delta_s(T)$
decreases. Similar to the CCO case in $x = 0.5$ LSCoO, CO is totally
uncoupled from SO, which simply follows the CO pattern.

\begin{figure}[!t]
\begin{center}
\includegraphics[width=0.7\linewidth]{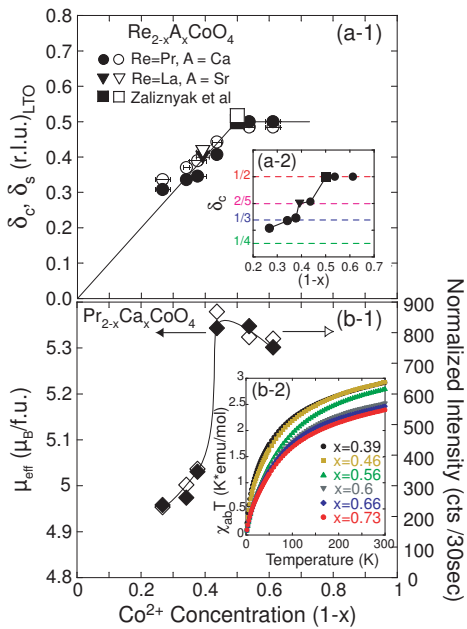}
\vspace{-0.1in}
\caption{(a) Diagram of the propargation vector for spin and charge
superstructures in PCCoO and LSCoO. Filled (open) symbols show the
in-plane modulation vector for charge (spin) order, $\delta_c =
\epsilon_c/2$ and $\delta_s$, respectively. Inset (a-2) is the
enlarged Yamada plot showing the lattice commensurability effect on
the position of CO peaks at ($2\delta_c,0,l$). (b) Concentration
dependence of $\mu_{\text{eff}}$ (filled symbols) and peak magnetic
neutron scattering intensity (open) in PCCoO. Inset (b-2) shows the
temperature dependencies of $T \cdot \chi_{ab}$ for selected $x$'s.
Lines are guides for the eye. \label{Fig3_PD-mu_int} }
\end{center}
\vspace{-0.4in}
\end{figure}

The doping dependencies of the incommensurabilities of charge and
spin order are summarized in Figure \ref{Fig3_PD-mu_int}(a). As
mentioned before, for $x \ge 0.5$ both roughly follow the $\sim 1-x$
trend, although $\delta_s(x)$ is always somewhat larger than
$\delta_c(x)$, corresponding to shorter average SO modulation
period. This again is consistent with the picture where
shorter-periodic polaron superlattices have better developed spin
correlations. Detailed investigation of the CO peak position, Fig.
\ref{Fig3_PD-mu_int}(a-2), shows lattice commensurability effect on
CO, indicating that most observed CO peaks can be assigned to a
mixture of two nearest commensurate superlattices
\cite{Yoshizawa_PRB2000_LSNO_stripe_step,
Kajimoto_PRB2003_LSNiO_stripe_PD}.

Finally, we comment on an apparent doping-induced Co spin state
change, which is manifested by a decrease of $\mu_{eff}(x)$ and in
PCCoO occurs at $x \gtrsim 0.56$, Figure \ref{Fig0_xdep}(c).
Comparison of Figure \ref{Fig3_PD-mu_int} (a) and (b) suggests that
other than decrease of the SO peak magnitude coincident with that of
$\mu_{eff}$ it does not have any significant effect on charge and
spin ordering.

In summary, we have discovered doping-dependent charge and spin
modulations in PCCoO and LSCoO, whose periods vary roughly linearly
with doping, $\delta_s \sim \epsilon_c/2 \sim 1-x$, existing for $x
>0.5$. They border the checkerboard CO phase at $x \lesssim 0.5$.
In cobaltites CO occurs in a phase where electrons are strongly
localized and can therefore be understood as a correlated polaron
glass with nanoscale patches of commensurate CO superlattices, whose
long-range coherence is frustrated by the charge neutrality
requirement. Antiferromagnetic SO correlations between the nearest
"undoped" Co$^{2+}$ sites develop at temperatures more than 100
times smaller than CO and do not affect it. Similar CO (but not SO)
was found in layered manganites, where it also emerges from the
state with thermally activated transport
\cite{Bao_SolStCom1996,Larochelle_PRB2005}. There, though, polaron
correlations are associated with the orbital order and
double-exchange physics, which are absent here. CO and SO in
cobaltites are most likely governed by the lattice electrostatics
and superexchange. However, in many respects they have quite similar
appearance to CO/SO in cuprates and nickelates, showing ubiquity of
this type of pattern.

We thank T. J. Sato and K. Hirota for help with experiments. This
work was partly supported by Grant-In-Aids for Scientific Research
(C) (No. 16540307) from the Ministry of Education, Culture, Sports,
Science, and Technology, Japan, and by the US DOE under the Contract
DE-\-AC02-\-98CH10886.


\end{document}